\documentclass[twocolumn]{aastex63}

\usepackage[utf8]{inputenc}
\usepackage{enumitem}
\usepackage{amsmath}
\usepackage{graphicx}
\usepackage{hyperref}
\usepackage{color}

\usepackage{rotating}

\usepackage{booktabs}
\usepackage{multirow}

\hypersetup{colorlinks=true, citecolor=cyan, urlcolor = blue, linkcolor = blue}

\shorttitle{Source Confusion for PTAs}
\shortauthors{Witt et al.}

\begin{document}
\title{Lost in the Signal Model: Confusion Between Deterministic and Stochastic Gravitational Wave Signals in Pulsar Timing Array Data}

\correspondingauthor{Caitlin A. Witt$^{\color{magenta}\S}$}
\email{wittc@wfu.edu}

\author[0000-0002-6020-9274]{Caitlin A. Witt$^{\color{magenta}\S}$}
\affiliation{Department of Physics, Wake Forest University, Winston-Salem NC 27105, USA}
\affiliation{Center for Interdisciplinary Exploration and Research in Astrophysics (CIERA), Northwestern University, Evanston, IL 60208}
\affiliation{Adler Planetarium, 1300 S. DuSable Lake Shore Dr., Chicago, IL 60605, USA}

\author[0000-0002-8826-1285]{Nihan S. Pol}
\affiliation{Department of Physics and Astronomy, Texas Tech University, Lubbock, TX 79409-1051, USA}

\begin{abstract}
With growing evidence for the presence of a gravitational wave background (GWB) in pulsar timing array (PTA) data, it is imperative to understand how this signal will act as a foreground for other GW signals, particularly those from individual supermassive black hole binaries (SMBHBs), which are predicted to be detected in the next decade. For PTAs, since these two types of signals are long-timescale persistent signals, a signal from an SMBHB system may already be present in the latest PTA datasets. In this work, we analyze how the presence of this signal could affect our detection statistics for the GWB, and how the presence of the GWB can affect our search for the SMBHB system. We find that individual SMBHB signals can significantly skew the recovery of the GWB amplitude and spectral index. However, when the SMBHB signals are recovered consistently, we are again able to recover the GWB power-law parameters, highlighting the importance of joint searches in future PTA searches for GWs. We also explore the rate of improvement of parameter estimation uncertainty, giving insight to the types of signals most likely to be identified well enough to enable multimessenger analyses.
\end{abstract}


\section{Introduction}


Supermassive black hole binaries (SMBHBs) are predicted to form in the aftermath of galaxy mergers \citep{Begelman1980}, and through long-term evolution, reach milli-parsec separations where they will be strong emitters of gravitational waves (GWs). At these small separations, direct observational confirmation has yet to occur, but numerous candidates have been proposed \citep{Sudou2003, Graham2015, charisi2016, panstarr}. As hierarchical galaxy mergers are thought to be common throughout the universe in standard cosmology, SMBHBs are theorized to exist as a population across cosmic time \citep{sesana_04}.
As the GWs emitted from these binaries combine into a superposition of signals, they will form a stochastic GW background (GWB) whose properties are related to the population of GW progenitors producing the signal. For an SMBHB-produced background, the superposition is predicted to form a power-law characteristic strain spectrum of the form $h_c \propto A_{\rm GWB} \left(f/f_{\rm yr}\right)^{-2/3}$ for purely GW-driven evolving systems \citep{RajagopalRomani95, JaffeBacker03, sesana08}. 

Such a background would primarily be detectable at nanohertz frequencies, and is therefore a primary source of interest for pulsar timing arrays \citep[PTAs,][]{PTAs, ptas2}. By measuring nanosecond-precision delays in the times-of-arrival (TOAs) from dozens of millisecond pulsars (MSPs) around the Galaxy, PTAs form a galactic-scale GW detector suitable for searching for the low-frequency GWB and other GWs in the nanohertz frequency range.

After decades of effort, PTA datasets recently published evidence for the presence of a GWB in their latest datasets. 
This evidence, to varying degrees, was reported by The North American Nanohertz Observatory for Gravitational waves \citep[NANOGrav;][]{nanograv}, the European Pulsar Timing Array \citep[EPTA;][]{EPTA}, the Parkes Pulsar Timing Array \citep[PPTA;][]{ppta}, and the Chinese Pulsar Timing Array \citep[CPTA;][]{cpta}.
While it is expected that the most likely source for a GWB in the nanohertz regime is a population of inspiraling SMBHBs \citep{review}, current analyses cannot yet identify if the progenitor of the GWB signal is indeed SMBHBs \citep{nano15astro} or another alternative source \citep{nano15new}.

If the origin of the GWB is indeed a population of SMBHBs \citep{sesana_04}, we would expect that one of the next possible GW sources that PTAs may uncover are ``continuous waves'' (CWs) from individual SMBHBs that are able to stand out above the GWB signal. In such a scenario, a joint analysis of the CW and GWB signals will be critical for accurate results \citep{mdc}, and such analyses have already been implemented for CW searches to place upper limits \citep{12p5cw, nano15cw, ipta2cw}. 
However, it is yet to be determined quantitatively how this CW signal will manifest itself in PTA datasets that already contain a GWB signal and, in particular, whether the two signals could interfere to produce biases in the inference of either of them.


This interplay between the signals may potentially already be affecting the GW searches in the latest datasets. Both the NANOGrav 15 yr and EPTA DR2 datasets showed marginal evidence for a CW signal at $f \approx 4$nHz when compared against a model that contains only a common, uncorrelated red noise (CURN) process among the pulsars in the PTA. However, the correct model for a GWB includes the characteristic GR-predicted quadrupolar Hellings \& Downs (HD) correlations \citep{hellings-downs} in the signal measured by the pulsars in the PTAs. On inclusion of these HD correlations in the GWB model\footnote{Note the CURN model places no requirement on inter-pulsar correlations}, the evidence for the CW signal in the data vanishes \citep{nano15cw}. However, an inclusion of the CW signal in the GWB model also works towards completely removing the HD correlations from the data \citep{nano15cw}. 

It is impossible to determine based only on the real datasets whether this behavior could be indicative of the emergence of a real CW signal and thus requires a dedicated investigation into these effects via extensive simulations. A first investigation of this interplay was performed in \citet{ferranti}, where they produced seven simulated datasets that replicated the real EPTA DR2Full dataset. Of these datasets, only 2 contain a GWB and a CW signal, with the injected noise and all CW parameters, except for frequency, being identical across these datasets. This minimal set of simulations allowed \citet{ferranti} to deeply investigate the effect of the presence of both signals on the parameter estimation. 

However, a lack of variation in the noise between datasets and CW parameters leaves the variance associated with their conclusions unclear. We complement their work by significantly increasing the variety in the injected CW sources, while also allowing the noise realizations to vary between datasets.
We base our simulations on the NANOGrav 15 yr dataset, which contains 67 pulsars timed across a maximum baseline of 16 yrs. This allows us to test the interplay between these two types of signals in a PTA that contains a significantly larger number of pulsars, which corresponds to better sensitivity to CW signals.
Additionally, across 3 choices for each of the CW chirp mass, frequency, and luminosity distance, and 100 noise realizations, we present an ensemble analysis of 8100 simulated datasets and exhaustively cover the effects of the presence of signals from both individual SMBHBs, as well as the entire GWB.





The remainder of our paper is organized as follows. In \autoref{sec:methods}, we describe our simulation and GW search methods. In \autoref{sec:results}, we present the results of our analysis. In \autoref{sec:discussion}, we discuss their implications for future multi-source GW searches by PTAs, and in \autoref{sec:conclusion}, we present our conclusions.

\section{Methods}\label{sec:methods}
\subsection{Simulations}
We begin with the NANOGrav 15-year dataset \citep{nano15timing} and generate 100 realizations of white noise, red noise, and GWB (where applicable).
These are created with the same simulation scheme as \citet{astro4cast}, where pulse time-of-arrival (TOA) timestamps and uncertainties are epoch-averaged to drastically reduce computation time, while still maintaining an accurate portrayal of the data set's sensitivity to GWs. These data sets each contain white and red noise based on the values measured in the 15-year data set.

\begin{figure}
    \centering
    \includegraphics[width = 1\columnwidth]{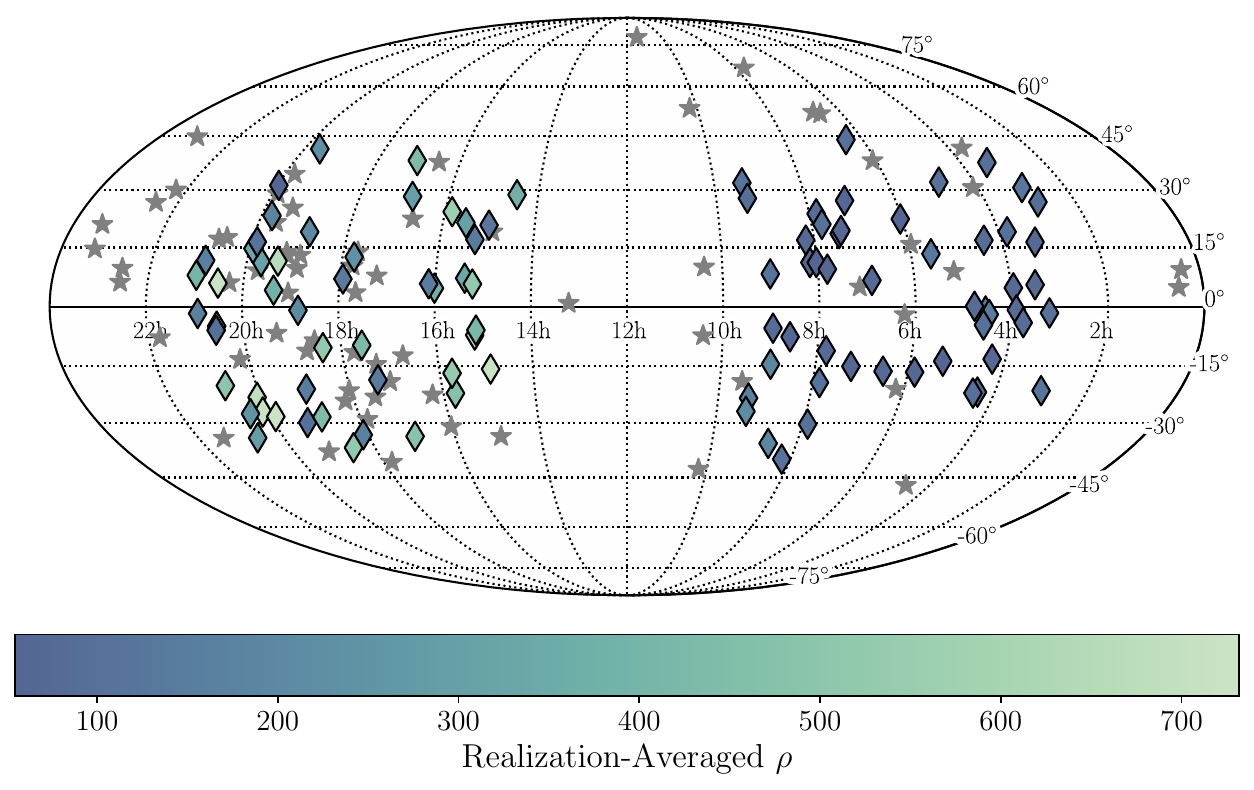}
    \caption{Sky locations of the pulsars in NANOGrav's 15-year data set (grey stars) and our injected SMBHB signals (diamonds). The color axis represents the log-likelihood ratio of each binary, averaged across each combination of injected parameters.}
    \label{fig:sim_map}
\end{figure}

Next, we inject one of three possibilities for the GWB into each data set: no GWB, where there remains only intrinsic pulsar noise, a common uncorrelated red noise (CURN) process, which contains power-law red noise common to each pulsar, and a full GWB process which contains power-law red noise common to each pulsar with Hellings-Downs \citep{hellings-downs} inter-pulsar correlations. Both the CURN and GWB data sets have amplitudes matching those recovered in the 15-year data set in \citet{nano15gwb}, $A_{\rm GWB} = 2.4\times 10^{-15}$ for a GWB model assuming an SMBHB origin with the power-law spectral index in timing residual units $\gamma_{\rm GWB} = 13/3$.

Finally, in each realization we inject a CW signal from an SMBHB. The SMBHBs are randomly distributed across two large areas of the sky generally corresponding to the PTA's most and least sensitive areas \citep{nano15cw}. This is done for three values of the SMBHB chirp mass $\mathcal{M}$, GW frequency $f_{{GW}}$, and luminosity distance $d_L$, resulting in 27 combinations listed in \autoref{tab:injected}, and therefore, 27 strain amplitude values, where the strain amplitude $h_0$ is
\begin{equation}\label{eq:strain}
    h_0 = \frac{2 {\mathcal{M}^{5/3}}(\pi f_{{\rm GW}})^{2/3}}{{d_L}},  
\end{equation}
in units where $G = c = 1$. Angular parameters (polarization angle $\psi$, initial phase $\Phi_0$, and orbital inclination $\iota$) are chosen at random for each realization. Each CW injection uses a full evolution model, where the orbital phase and angular frequency $\omega_0 = \pi f_{\rm GW}$ evolve as
\begin{align}
    \Phi(t)&=\Phi_{0}+\frac{1}{32} \mathcal{M}^{-5 / 3}\left[\omega_{0}^{-5 / 3}-\omega(t)^{-5 / 3}\right], \\ \omega(t)&=\omega_{0}\left(1-\frac{256}{5} \mathcal{M}^{5 / 3} \omega_{0}^{8 / 3} t\right)^{-3 / 8}.
\end{align}
We also include a full pulsar-term model to account for the GW-driven evolution of the binary during the light-travel time between the Earth and each pulsar, where the difference between the signal $s$ at the Earth and at the pulsar $\Delta s$ is calculated as 
\begin{equation}
    \Delta s_{+, \times}(t)=s_{+, \times}\left(t_{p}\right)-s_{+, \times}(t),
\end{equation}
where 
\begin{equation}
    t_{p}=t-L(1+\hat{\Omega} \cdot \hat{p})
\end{equation}
and $L$ is the distance to the pulsar. Therefore, the full signal can be written as 
\begin{equation}
    s(t, \hat{\Omega})=F^{+}(\hat{\Omega}) \Delta s_{+}(t)+F^{ \times}(\hat{\Omega}) \Delta s_{ \times}(t),
\end{equation}
where $F^{+,\times}$ are the GW antenna pattern functions \citep{sesanavecchio, ellis2012, taylor2016}, and 
\begin{align} 
\begin{split}
    s_{+}(t)=& \frac{\mathcal{M}^{5 / 3}}{d_{L} \omega(t)^{1 / 3}}\left[-\sin 2 \Phi(t)\left(1+\cos ^{2} i\right) \cos 2 \psi\right.\\ &-2 \cos 2 \Phi(t) \cos i \sin 2 \psi], \\ s_{ \times}(t)=& \frac{\mathcal{M}^{5 / 3}}{d_{L} \omega(t)^{1 / 3}}\left[-\sin 2 \Phi(t)\left(1+\cos ^{2} i\right) \sin 2 \psi\right.\\ &+2 \cos 2 \Phi(t) \cos i \cos 2 \psi].
\end{split}
\end{align}

The locations of the SMBHB injections relative to the pulsars in the 15-year data set are plotted in \autoref{fig:sim_map}. The points are colorized by each binary's calculated log-likelihood ratio, 
\begin{equation}\label{eq:loglike}
    \rho = \sqrt{2(\mathcal{\mathrm{log}L}_{\rm signal} - \mathrm{log}\mathcal{L}_{\rm noise})},
\end{equation}
averaged across each of the 27 combinations of injected source parameters, where $\mathrm{log}\mathcal{L}$ is the PTA log-likelihood originally defined in \citep{vHlevin}.
Across these axes, we have constructed 8100 PTA data sets for analysis.

\begin{table}[]
    \centering
    \begin{tabular}{ccc}
    \hline
        $\mathcal{M}$ ($M_\odot$) & $f_{GW}$ (Hz) & $d_L$ (Mpc)\\
        \hline
        \hline
        $1\times10^9$ & $3.87\times10^{-9}$ & 25 \\
        $5\times10^9$ & $1.35\times10^{-8}$ & 50 \\
        $1\times10^{10}$ & $1.55\times10^{-8}$ & 75 \\
        \hline
    \end{tabular}
    \caption{Injected SMBHB values.}
    \label{tab:injected}
\end{table}

\subsection{GW Searches} \label{subsec:gw_searches}
For each GW analysis, we use standard NANOGrav pipelines to replicate searches on real data; therefore, we will give a brief overview of the relevant components, and refer to \citet{nano15gwb} for more details.

The pulsar TOAs can expressed via separate components as
\begin{equation}
    \delta t = M \epsilon + n_{\mathrm{white}} + n_{\mathrm{red}} + s,
\end{equation}
where $M$ is the design matrix, which describes the linearized timing model, and $\epsilon$ is a vector of the timing model parameter offsets. As in analyses in real data, this term allows the timing model parameters of each pulsar to be adjusted in accordance with the presence of any modeled signals. The variables $n_{\mathrm{white}}$ and $n_{\mathrm{red}}$ refer to vectors describing the pulsar white and red noise, respectively, and $s$ is a vector of GW-induced signal present in the residuals. 

In our analyses, the white noise parameters are held fixed to their injected values. The power spectral density of each pulsar's red noise is fit to a power-law model as in \citet{shannon2010, lam2017}, where
\begin{equation}\label{eq:rn}
    P(f)=\frac{A_{\mathrm{red}}^{2}}{12\pi^2}\left(\frac{f}{f_{\mathrm{yr}}}\right)^{-\gamma_{\mathrm{red}}}~\mathrm{yr}^3.
\end{equation}
Here, $f_{\mathrm{yr}}\equiv 1/(1~\mathrm{year})$, $A_{\mathrm{red}}$ is the red noise amplitude, and $\gamma_{\mathrm{red}}$ is the power law spectral index. 
Each pulsar's noise spectrum is divided into 30 linearly spaced bins, ranging from $1/T_{\mathrm{obs}}$ to $30/T_{\mathrm{span}}$, where $T_{\mathrm{span}}$ is the total observation baseline for the PTA.
The prior on $A_{\mathrm{red}}$ is log-uniform in the range $[-18,-11]$, while the prior on $\gamma$ is uniform in the range $[0,7]$. 

On each data set created, we first conducted two main searches; one for a CURN process, and one for a CURN+CW model. We used the \texttt{enterprise} \citep{enterprise} package to construct the priors and evaluate the likelihood \citep{ellis2012}, as it has been thoroughly tested for PTA GW analyses \citep{codereview}. The Markov Chain Monte Carlo (MCMC) sampler package \texttt{PTMCMCSampler} \citep{ptmcmc} was used to explore the parameter space and construct our posterior distributions for analysis. To sample the likelihood space, we used identical parameter groupings and jump proposals as real, state-of-the-art \texttt{enterprise}-based CW analyses.

We also conduct an ``optimal statistic" \citep{anholm2009, 5yrgwb, chamberlain2015} analysis for each data set to analyze the correlations recovered from each dataset. The metric $\hat{A}^2$ measures the amplitude of the GWB corresponding to the sum of correlations between pulsar pairs, and is also weighted by intrinsic pulsar noise and conditioned on the assumed spectral index of the GWB. This procedure allows us to calculate a signal-to-noise ratio of spatial correlations from any template (HD for GW-induced quadrupolar correlations, or monopolar or dipolar correlations from other sources) \citep{chamberlain2015}. Here, we use the multiple-correlation optimal statistic \citep[MCOS,][]{mcos}, which simultaneously fits the inter-pulsar correlation coefficients with multiple correlation patterns (Hellings-Downs, monopolar, and dipolar).
We also test the response of the noise-marginalized optimal statistic \citep[NMOS,][]{nmos}, which uses posterior samples from a Bayesian analysis to marginalize over intrinsic pulsar red noise and calculate distributions of $\hat{A}^2$ and $S/N_{\rm HD}$
 and avoid contamination from covariances induced when pulsars contain intrinsic red noise, as well as the noise-marginalized multiple-correlation optimal statistic \citep[NMMCOS,][]{mcos}, which combines the two methods. 


The above methods model the GWB using a power-law template. However, we are also interested in investigating the effect of the presence of an unmodeled CW signal on the spectral recovery of any GWB signal. Since a simple power-law model may subsume these effects into variations in the amplitude and spectral index, we also model our CURN process with a Bayesian periodogram, i.e. a ``free-spectrum'', where the power in each Fourier bin is modeled independently and simultaneously. This allows us to parametrize the the spectral content of the recovered GW signal (modeled as a CURN process) as the variance of the Fourier components at each frequency, as in \citet{12p5_gwb} and \citet{nano15gwb}.

These methods together reflect the analyses used on real PTA data sets for production-level significance and parameter estimation \citep{nano15gwb}. Therefore, our results are a critical preview of the impact our first-line analyses will face as CWs become detectable above the GWB in PTA data sets.
\section{Results}\label{sec:results}

\subsection{Ensemble Analyses}\label{ss:pop}

We begin by passing all of the 8100 simulated datasets through the GWB detection pipeline described in Sec.~\ref{subsec:gw_searches}. We purposely exclude a CW signal from the model so that we can investigate the effect of an unmodeled CW on the recovery of the GWB.

\subsubsection{Parameter Estimation}\label{ss:pe}

\begin{figure}
    \centering
    \includegraphics[width=0.99\columnwidth]{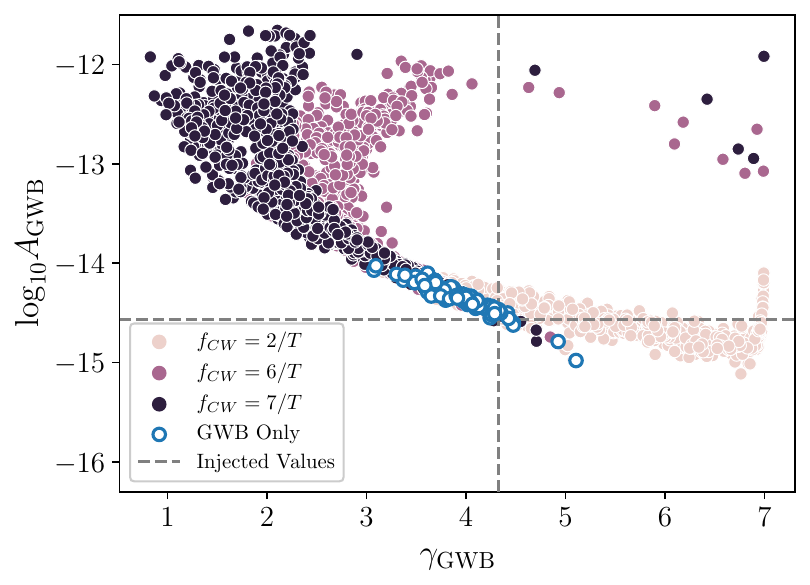}
    \caption{Recovered GWB parameters for a power law fit ($\gamma_{\rm GWB}$ and $\mathrm{log}_{10}A_{\rm GWB}$) to datasets containing both a GWB and a CW. The difference from the injected values (gray dashed lines) varies strongly as a function of the CW frequency $f_{CW}$ (purple hues). As a comparison, the recovered parameters for GWB-only data sets are shown (open blue points).}
    \label{fig:skew_gwb}
\end{figure}

\begin{figure*}
    \centering
    \includegraphics[width=0.99\linewidth]{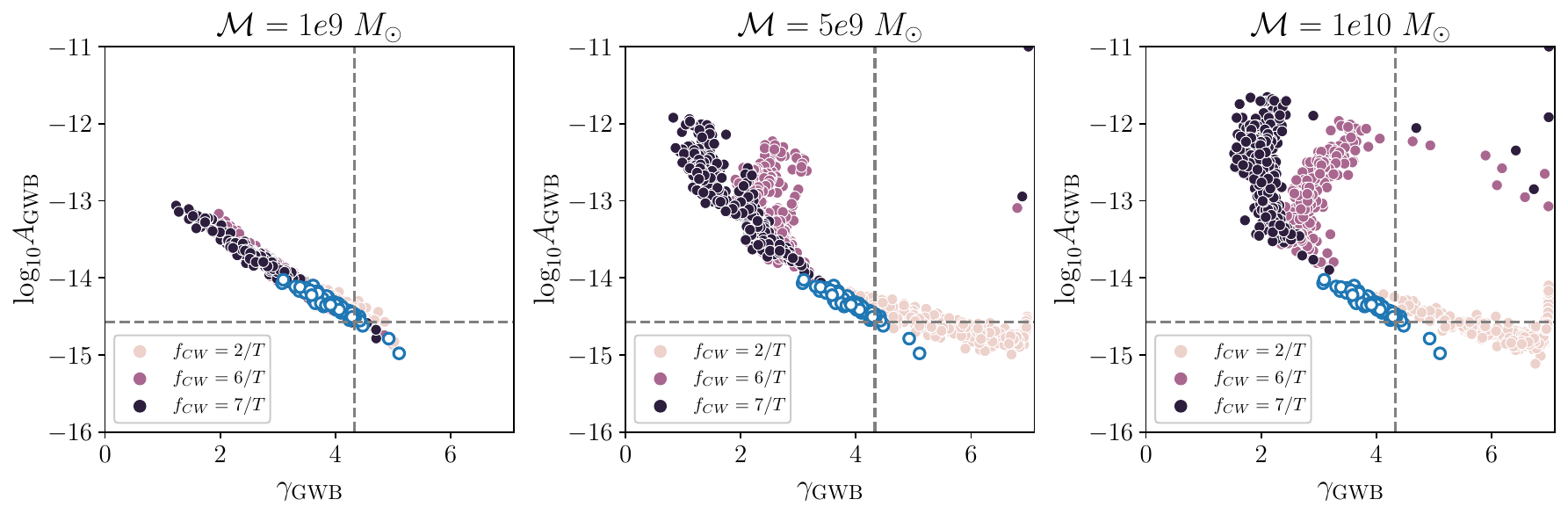}
    \caption{Recovered GWB parameters for a power law fit ($\gamma_{\rm GWB}$ and $\mathrm{log}_{10}A_{\rm GWB}$) to datasets containing both a GWB and a CW, separated by the chirp mass $\mathcal{M}$ of the injected signal. }
    \label{fig:mc}
\end{figure*}

As shown in \autoref{fig:skew_gwb}, we observe that the bias induced in a power-law GWB recovery varies dramatically depending on the frequency of the injected CW, and is significant compared to the spread of median posterior values of $A_{\rm GWB}$ and $\gamma_{\rm GWB}$ recovered in the same analysis for GWB--only data sets that do not contain a CW. At $f_{\rm GW} = 2/T$, the recovered CURN parameters are consistently of a similar strength ($A_{\rm GWB}$) but with a steeper spectral index. However, at higher frequencies, the effect is reversed, and the recovered CURN process is consistently stronger and with a shallower spectral index. We also observe that this scatter is larger for CWs injected with $\phi < 12$h, where our PTA is generally less sensitive.

The biases described above are also modulated by the chirp mass (and therefore, strain) of the injected CW source.
As shown in \autoref{fig:mc}, for moderate chirp masses, while there is still an induced bias, it follows the same trend seen in the range of recovered $A_{\rm GWB}$ and $\gamma_{\rm GWB}$ values in the base GWB-only data that does not contain an injected CW (\autoref{fig:mc}, left panel). 
For high chirp mass sources, we see a significant amount of structure in the induced bias that is dependent on the frequency of the injected signal (middle and right panels). 

\begin{figure*}
    \centering
    \includegraphics[width=0.99\linewidth]{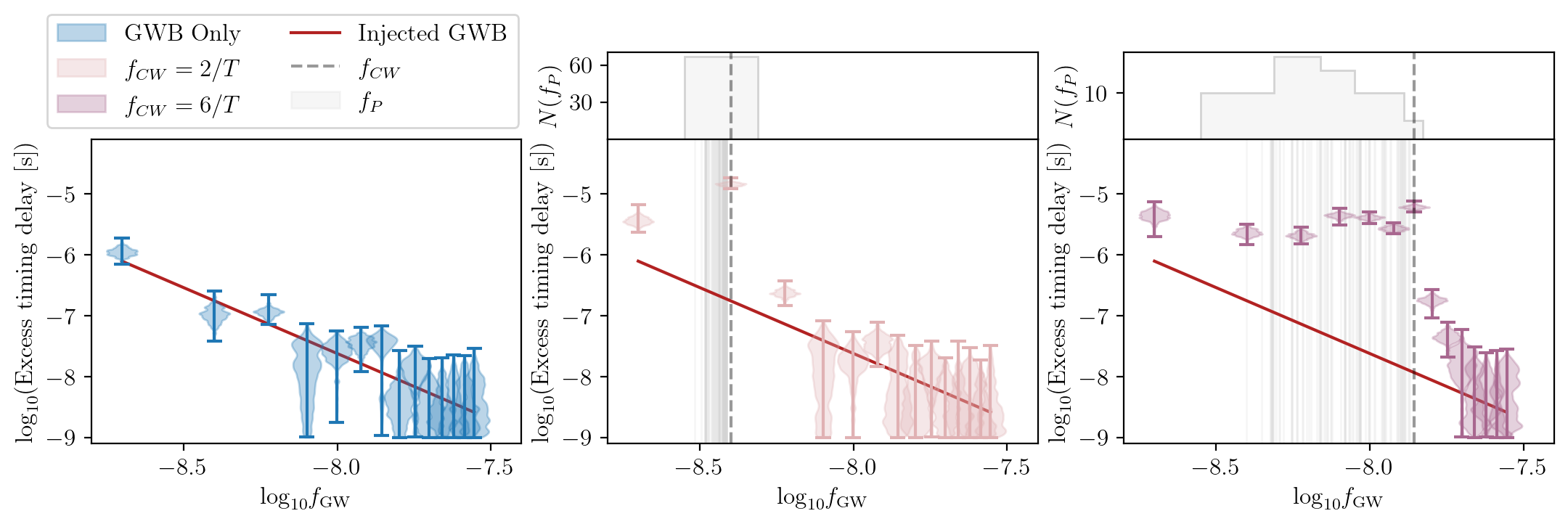}
    \caption{Bayesian ``free-spectrum" analyses showing posteriors (violins) of variance at frequencies $i/T$ in one realization of our data set containing either a GWB (blue) or a GWB+CW (peach and purple). These illustrate the recovery of injected power-law GWB (red line) and CW (gray dashed line). The CW frequencies seen in each pulsar (the "pulsar-term frequencies") are shown in light gray.}
    \label{fig:freespec}
\end{figure*}

To investigate the root cause of this effect, we calculate the Bayesian spectrum of the CURN for a subset of our simulated data sets.  
In \autoref{fig:freespec}, we can see the spectrum in one realization of our data set with no CW compared to a low- and high-frequency CW. In the GWB-only case, the measured spectrum accurately traces the injected power-law GWB signal. However, in data sets with a CW, it is readily apparent why the excess power induced by the CW significantly skews a power-law fit. In the low-frequency CW case, the signal significantly elevates the residual delay in the frequency bin containing the CW signal, which can no longer be accurately fit with a single power-law. In the high-frequency case, this effect is exacerbated, and we now observe excess power across multiple frequency bins. We trace this effect to the pulsar term frequencies included in the signal, which for this high Earth-term frequency, span the frequency bins marked in \autoref{fig:freespec}, which correspond to the injected CW parameters via 
\begin{equation}
    f_p =  f_{GW} \times (1 - \frac{256}{5} \mathcal{M}^{5/3} (\pi f_{GW})^{8/3} (t_p))^{-3/8},
\end{equation}
where $t_p = d_p/c \times (1-\mathrm{cos}\mu)$, and $\mu$ is the sky angle between the pulsar and SMBHB \citep{corbincornish}. This contextualizes the $\mathcal{M}$ correlations observed in \autoref{fig:mc}, as with higher values of $\mathcal{M}$, there is a more significant amount of evolution in the signal between the Earth- and pulsar-terms, resulting in excess power spread over more frequency bins, subsequently altering the shape of the recovered CURN spectrum when the CW is not accounted for.

\begin{figure*}
    \centering
    \includegraphics[width=0.99\linewidth]{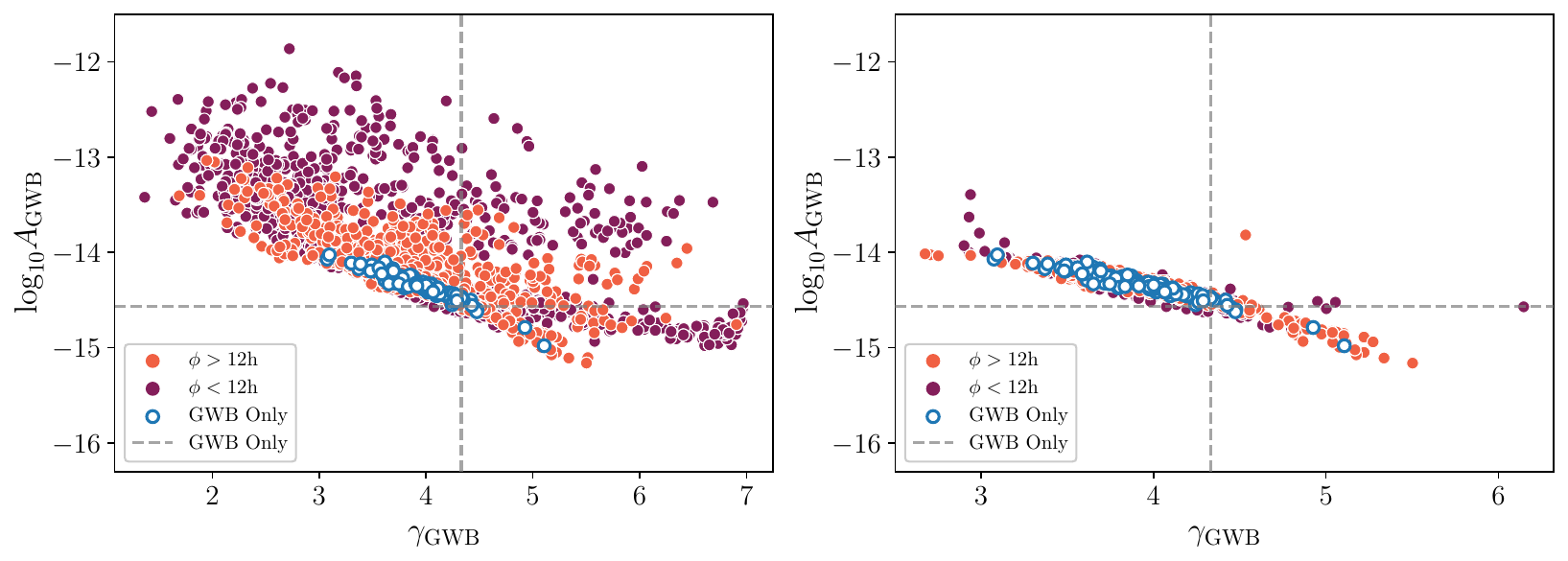}
    \caption{When jointly modeling CWs and a CURN process, the clear $\mathcal{M}$ and $f_{\rm GW}$ correlations disappear (left panel), but there is still significant skew between the recovered GWB parameters when there is a CW injected (orange and purple points) versus without (blue circles). This bias is larger for CWs injected at sky locations with $\phi < 12$h. If the CW parameters are consistent with the injected values (right panel), there is no longer a significant difference in our parameter estimation as compared to the GWB-only data sets. }
    \label{fig:cw_model}
\end{figure*}

To remedy this bias in our parameter estimation capabilities, we conducted CW+CURN analyses on the same simulated data sets. However, as can be seen in the left panel of \autoref{fig:cw_model}, while this does remove the evolution-induced structure within the biased GWB parameters, there is still significantly larger scatter in the GWB recovery compared to the datasets with no injected CWs.
This scatter is smaller for CWs injected with $\phi > 12$h, where our PTA is generally more sensitive.
We then reduced our sample to only those simulations where the key CW parameters ($h_0$, $\mathcal{M}$, $f_{\rm GW}$, and the source sky location) were recovered ``consistently'', which we identify as when the injected value falling within the 1-$\sigma$ confidence interval of the 1-d posterior. This sub-sample of our simulations (right panel of \autoref{fig:cw_model}) no longer displays the significant scatter in the jointly-modeled GWB parameters, and instead recovers $\mathrm{log}_{10}A_{\rm GWB}$ and $\gamma_{\rm GWB}$ to the same level of accuracy as the simulations without CWs. 

We quantify this accuracy by calculating the average Mahalanobis distance \citep{mahalanobis} of each population from the injected values in 2-D $\mathrm{log}_{10}A_{\rm GWB} $$-$$\gamma_{\rm GWB}$ space, which is scaled by the average number of standard deviations present in the 2-D data set. 
In GWB-only data sets, our recovered $\mathrm{log}_{10}A_{\rm GWB} $ and $\gamma_{\rm GWB}$ span an average Mahalanobis distance of
$ 0.53 \pm ^ {0.44} _ {0.44} $ from the injected value. 
In datasets with a CW where the CW is included in the signal model, the average Mahalanobis distance increases to 
$ 2.52 \pm ^ {1.21} _ {1.75} $. If we limit our populations to only simulations where the CW parameters were consistent with the , the average Mahalanobis distance is only, 
$ 1.29 \pm ^ {2.12} _ {1.1} $, which is much more consistent with the GWB-only data set. 
These values are also consistent with the uncertainty in the recovery in NANOGrav's 15-year data set
and follow a similar correlation between the two parameters.



We note that only about forty percent of simulations fell within this ``consistently recovered CW" category. 
While the recovery across our 100 realizations is correct on average, and there is no uniform bias suggesting a mis-modeling, careful analyses will be needed in the future as the first CW is detected and posteriors begin to narrow. 

\subsection{CURN Significance}

\begin{figure}
    \centering
    \includegraphics[width=1 \columnwidth]{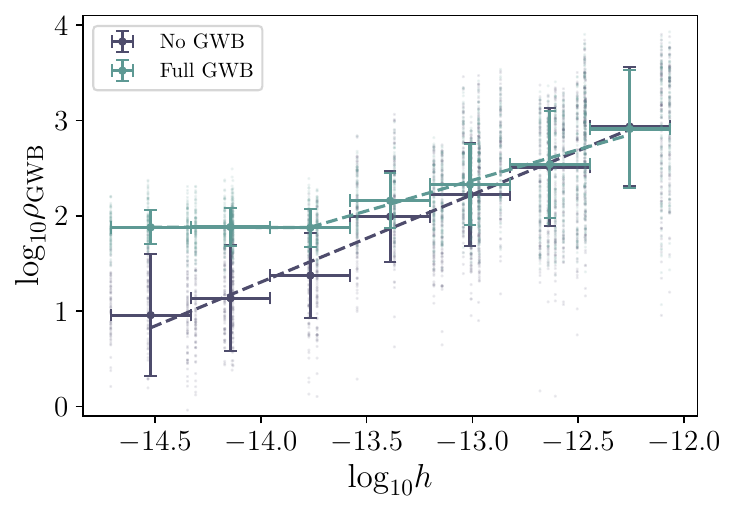}
    \caption{As expected, increasing the strain amplitude of a CW results in a stronger resulting value of $\rho_{GWB}$ due to similarities between the two signals (blue points). However, when a true GWB is also present, the background saturates the $\rho_{\rm GWB}$ calculation at low injected CW strains, producing GWB results as expected without a CW present.}
    \label{fig:scaling}
\end{figure}

To evaluate the significance of the CURN signal recovered in our analysis, we utilize a log-likelihood ratio $\rho$ defined in \autoref{eq:loglike}. 
We observed a significant increase in the realization-averaged value of $\mathrm{log}_{10}\rho$ in nearly all data sets containing a CW, whether or not a GWB was truly present. We observe a correlation between the injected strain of the CW and our measured value of $\mathrm{log}_{10}\rho$ (\autoref{fig:scaling}) that we fit as $\rho \propto h^{0.9\pm 0.3}$.
However, in data sets that do contain a GWB, the power-law correlation breaks for low values of $h_{\rm CW}$ where the GWB is stronger than the injected CW. Below the break strain (approximately $\mathrm{log}_{10}h = -13.5$), the fit is flat with $\rho_{\rm GWB} \propto h^{0.0\pm 0.3}$, and above the break frequency the power-law fit is consistent with the CW-only data set, with $\rho_{\rm GWB} \propto h^{0.6\pm 0.3}$.
Therefore, we expect that as a CW emerges in our data, existing GWB significance metrics will be robust to low-strain CWs, but would break down for very strong signals. Similarly, we can infer that as the signal-to-noise ratio of a CW signal grows, we may observe similar behavior, but this must be tracked in future work, as discussed in \autoref{sec:discussion}.

\begin{figure*}
    \centering
    \includegraphics[width=1\linewidth]{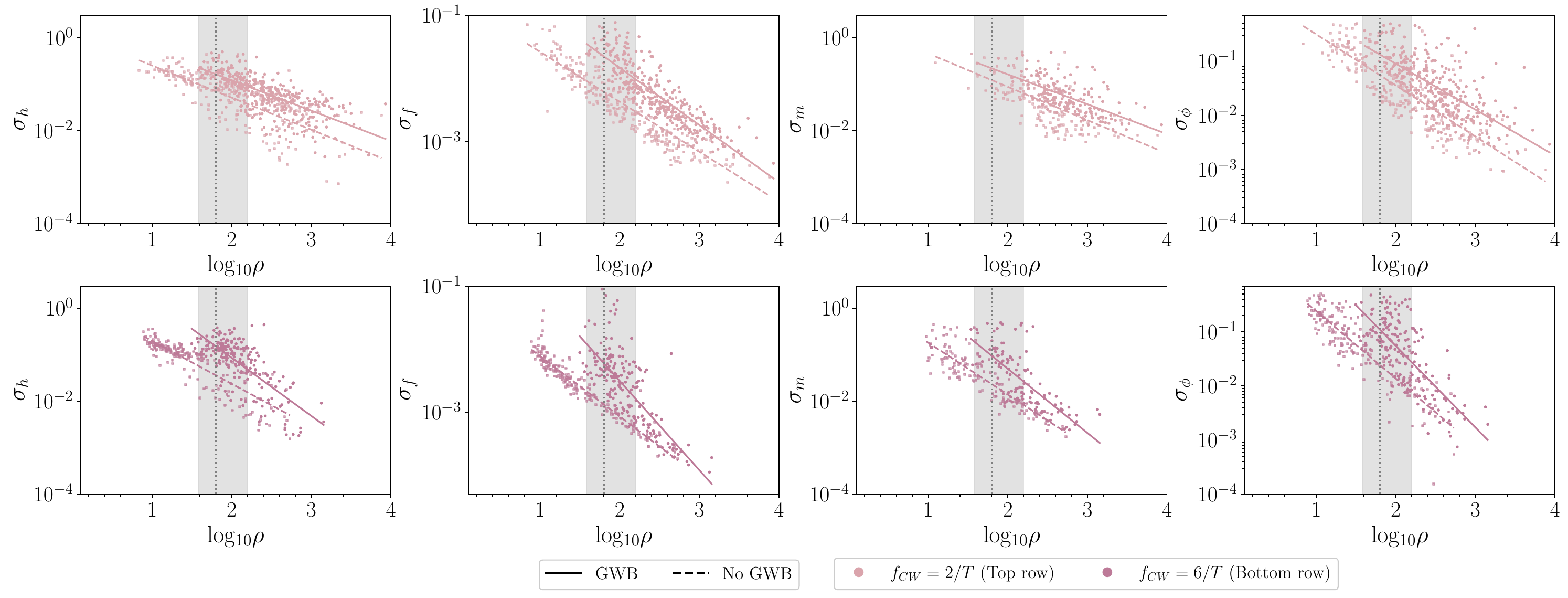}
    \caption{As the log-likelihood ratio of the GWB+CW model increases, the uncertainty on the parameter estimation of strain, CW frequency, chirp mass, and sky position (purple points) tightens according to a power law (purple curves). The slope of the uncertainty evolution is strongly dependent on if the GWB is present (solid curves) or not (dashed curves) as well as the frequency of the observed CW (color). }
    \label{fig:tracks}
\end{figure*}

We explore how parameter estimation will evolve as a function of the total model significance by modeling the 1-D one-sigma confidence interval as a function of $\mathrm{log_{10}}\rho$ for the GWB+CW model (\autoref{fig:tracks}). Generally, the one-sigma uncertainty of the global CW parameters strain, frequency, chirp mass, and sky location $\phi$ follow a
power-law, 
\begin{equation}
    \sigma(\rho) = \sigma_{\text{min}} \left( \frac{\rho}{\rho_{\text{min}}} \right)^{\alpha},
\end{equation}
where $\sigma_{\text{min}}$ and $\rho_{\text{min}}$ are reference values of the likelihood ratio and uncertainty (defined here at the minimum value of $\rho$ in each data set) and $\alpha$ is the power-law index. Here, we calculate the fit
in each case where the posterior does not return the prior (i.e. searches which constrain each of these parameters). The evolution of size of the confidence interval of each parameter is strongly dependent upon the frequency of the signal, and the presence of the GWB strongly affects the initial evolution of the confidence interval (dashed curves). However, for high values of $\mathrm{log_{10}} \rho$, where the CW is distinct from the underlying GWB, the confidence interval evolution tracks display consistent behavior among the four parameters. For the lowest CW frequency $f_{\rm GW} = 2/T$, the parameter constraints display a moderate tightening with a slope of $\alpha \approx -0.6 $ to $ -0.9$ for each parameter. As $f_{\rm GW}$ increases (and therefore, so does the number of cycles observed) so does the rate of tightening for each parameter, increasing to $\alpha \approx -1.5 $ to $ -1.7$.  A full table of the fit parameters is shown in \autoref{sec:appendix}.

\subsubsection{Inter-pulsar correlations}\label{ss:os}


\begin{figure}
    \centering
    \includegraphics[width=1\columnwidth]{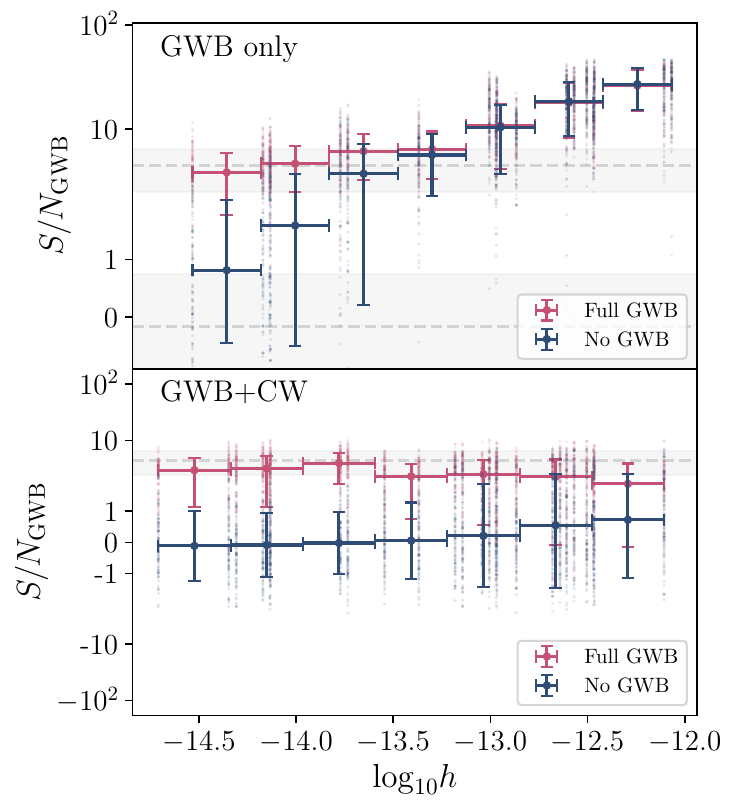}
    \caption{GWB signal-to-noise ratios calculated using the NMOS, based on analyses that only model a CURN process (top panel) versus those that also include a CW in the signal model (bottom panel). Neglecting to include the CW induces a significant bias in the MCOS S/N that is correlated to the strain of the CW. Modeling the CW removes the bias, and successfully recovers the original average HD S/N present in the data sets before the CW injection.}
    \label{fig:OS}
\end{figure}


The presence of spatial correlations consistent with the Hellings-Downs (HD) model is critical for confirming the presence of an astrophysical origin for a GWB. However, the presence of a CW could easily mask an uncorrelated red noise process and cause it to be mis-identified as an HD correlated GWB \citep{cornishsesana}. To ensure that PTAs will be able to continue to measure the significance of HD correlations as CWs emerge in future data sets, we investigate the recovered signal-to-noise ratio of HD correlations in our simulated data sets using the MCOS (\autoref{fig:OS}). We observed that when we calculate the S/N of HD correlations in data sets containing a CW (from analyses conducted with a CURN+CW model), there is a direct correlation between the injected CW strain and the recovered GWB S/N. If the data set also contains a GWB, the S/N that is indeed due to the GWB can be observed at low CW strains, but is saturated by the CW at high strain values. If we instead model both a CURN and a CW before calculating the MCOS, the measured S/N is constant across the entire range of injected strain values, and consistent with the average S/N present in the data sets before the CW injection. The S/N values for monopolar and dipolar correlations are consistent with zero across our 100 realizations and are not affected by the strain of the injected CW or if the CW parameters are accurately recovered. However, individual realizations do create S/N values above zero, presenting a clear need for careful modeling in real data.
We repeat these calculations with the NMOS and NMMCOS and find nearly identical results. However, the NMOS values for the analyses where the CW is modeled are systematically lower than those without, and the NMMCOS S/N values contain a significantly larger variance across the 100 realizations.








\section{Discussion and Future Prospects}\label{sec:discussion}

Through this work, we have explored the response of state-of-the-art PTA pipelines that search simultaneously for both a GWB and CW signal in the same dataset. We did this across a broad range of sky positions for the CW as well as 100 realizations of the the intrinsic pulsar red noise and GWB. The first conclusion that we can draw from the suite of analyses that we performed is that these joint-search pipelines are not guaranteed to accurately recover the parameters of both the injected GWB and CW. We emphasize that our analysis pipeline included both custom groupings and jump proposals for the parameters in the analysis, but despite this, only $\sim$30\% of our simulations were able to accurately recover both the GWB and CW signals injected into the dataset. This effect is most prominently displayed in \autoref{fig:cw_model}, where we see the effects of the mismodeling of the CW parameters on the recovered GWB parameters. 
We also see that injections with $\phi < 12$h show a larger fraction of inaccurate recoveries, which can be attributed to the lower density of pulsars in that part of the sky. 

Despite the difficulty in accurate CW parameter recovery, we highlight that the observed spectrum of the GWB could provide an intrinsic diagnostic to the accuracy of the recovered CW parameters. As shown in the second and third panels of \autoref{fig:freespec}, the CW signal manifests as a perturbation across one or more frequency bins, depending on the frequency of the CW source and the ``pulsar-term'' contributions to the signal. An accurate recovery of the CW signal should produce a power-law template for the GWB spectrum. However, if the spectrum of the GWB is expected to deviate from a simple power-law nature, as would happen at sufficiently low-frequencies for an SMBHB-produced GWB \citep{nano15astro} or if there are two underlying physical processes with different spectral indices contributing to the measured total GWB \citep{nano15new}, this verification method may not be reliable in and of itself.

The biggest current impediment to solving the problem of accurate GWB+CW parameter recovery is the 2--3 day run-time for each of these analyses with the current {\sc enterprise}-based pipelines. Thus, new software will be required that can significantly improve the run-time for these analyses. Fortunately, new software that can leverage GPU-acceleration are currently under development, including {\sc discovery} and {\sc prometheus} \citep{discovery, gundersen2026}. These new software packages will allow us to explore not just different sampling techniques, but also whether any modifications are required to the current analysis for improved parameter recovery. For example, The IPTA CW Mock Data Challenge \citep{mdc} showed that modeling white noise simultaneously with the CW signal was necessary to adequately detect the CW.
Additionally, it has been demonstrated that advanced noise models with custom per-pulsar Fourier bases can affect searches for CW signals and may also be required for accurate parameter recovery of CW signals \citep{15yr_CNM, ipta_dr2_cw}. Altogether, PTAs may also require a LISA-style Global Fit analysis pipeline where all components of the noise in the PTA are fit simultaneously with the GW signals of interest in order to claim a confident joint detection of a GWB and CW signal \citep{lisaglobal}.

In addition to parameter estimation of the GW signals, we also investigate how the presence of a CW signal can affect the measured HD S/N for the GWB as measured via a CURN and OS analysis. We show in \autoref{fig:OS} a direct correlation between the strain of the CW and the measured OS HD S/N in the absence of a GWB signal (upper panel) which saturates at low CW strains when a GWB signal is present in the dataset. Notably, we see that when a CW signal is modeled in addition to a CURN signal, the correlation with CW strain disappears with the recovered HD S/N corresponding to that of the injected GWB or noise, as shown in the lower panel of \autoref{fig:OS}. We can compare this behavior to that seen in the search for a CW signal in the NANOGrav 15 yr dataset \citep{nano15cw}. In that search, inclusion of a CW signal with the CURN process resulted in the measured optimal statistic HD S/N to reduce significantly and be consistent with zero. This matches our simulation set where we only had a CW injected in the dataset without a GWB. However, we caution against making the conclusion that this implies the real NANOGrav 15 yr dataset contains only a CW signal with no GWB. 
Our simulations only explore the interplay between these measured statistics in the case of a CW+GWB injection, rather than any other physical processes which may masquerade as a weak CW signal. Therefore, we maintain the possibility of noise fluctuations, either in the intrinsic pulsar noise or the GWB itself, masquerading as a weak CW signal in the real NANOGrav 15 yr dataset.

The results we present in this work also agree with those presented in other works while also expanding on them in several key areas.
We observe the same biases observed by \citet{ferranti} in the GWB parameter recovery when the CW signal is not modeled, however, unlike their limited (yet in-depth) analysis, we show that this bias depends on the chirp mass, frequency, and sky location of the CW source. However, unlike the conclusions of \citet{ferranti}, we demonstrate through our larger simulation suite that simply including the CW in the model is not guaranteed to accurately recover both of the signals that might be present in the data. We also explicitly show how a CW signal can affect the total measured spectrum for the GWB via the emergence of pulsar terms as unmodeled noise that gets attributed to the modeled common process. However, together, both of our analysis emphasize the need to carefully model the GWB and the CW in future PTA datasets.
However, \citet{becsy_pop} did not observe the biases that this work and \citet{ferranti} observed, due to the moderate strength of the signals injected. 
In future analyses, care must be taken with simulation efforts to replicate the astrophysical sources that PTAs will observe, as these discrete techniques converge on a consensus.  

Finally, in \autoref{fig:tracks}, we show scaling relations for the uncertainty on the recovered CW parameters as a function of the total S/N of the GWB+CW model. 
The log-likelihood ratio $\rho$ will increase with the strength of a CW, but can also be viewed as a proxy for signal-to-noise ratio, which will increase as our PTA becomes more sensitive. These parameter uncertainty scaling relations will prove critical as CWs become dominant over the GWB and PTAs begin to resolve individual sources, when identifying a host galaxy will be the ultimate goal. In particular, sky location and chirp mass constraints will immediately assist as multimessenger analyses aim to identify a host galaxy \citep{petrov_delphi}. However, the variance with frequency means that host identification prospects will vary heavily with the orbital period of the SMBHB, with binaries with higher orbital frequencies or orbital frequencies lying nearest to the most sensitive frequency of the PTA being the quickest to constrain, and therefore, the best targets for electromagnetic surveys.
Conversely, the total S/N of the GWB+CW model also depends on the dataset that was used in these analyses, allowing us to test the accuracy of recovered parameters across different datasets. In particular, it can allow us to predict the accuracy with which future datasets, such as those with the DSA telescope, would recover CW parameters. To do this, we would need a simulation of the TOA dataset from these facilities with an injected CW+GWB signal, and then directly calculating the total S/N statistic using the known signal values. This can then be combined with our scaling laws to estimate the uncertainty on the CW parameters without requiring a suite of computationally expensive and expansive injection-recovery analyses. Thus, such scaling laws can be useful in designing and optimizing future PTA surveys for detecting and characterizing CW sources in the presence of a GWB.
\section{Conclusions}\label{sec:conclusion}
Through the creation of an extensive simulation bank of PTA data sets and detailed Bayesian analyses, we have conducted an analysis of the biases induced on GWB analysis by CWs present in PTA data sets. 
We observed that strong CWs significantly bias the power-law parameter estimation of GWB recovery analyses, which is only alleviated when the CW signal is correctly identified. Through detailed spectral investigation, we discover that this is due to the presence of the pulsar-term of the CW signal, which spreads power across multiple GW frequency bins and subsequently biases the GWB spectrum recovered if the CW is not properly modeled. 

These biases extend to GWB significance measurements as well, as illustrated by log-likelihood ratio analyses for the GWB signal and the NMOS, both of which display a strong correlation with the strain of the injected GWB. Fortunately, modeling the CW alongside the GWB removes this correlation in the NMOS analyses, but significant work remains for PTA collaborations to ensure robust measurements of the GWB's significance and spectrum









\section{Acknowledgments}\label{sec:acks}
\textit{Author contributions:}
C.A.W. and N.S.P. originated the idea. 
C.A.W. created the simulated data sets, conducted GWB analyses, created figures, and drafted the manuscript. 
N.S.P created the data simulation techniques, conducted CW analyses, and provided commentary on the draft. 

\textit{Acknowledgments.}
C.A.W. acknowledges start-up funds from Wake Forest University and support from CIERA, the Adler Planetarium, and the Brinson Foundation through a CIERA-Adler postdoctoral fellowship. N.S.P acknowledges start-up funds from Texas Tech University. C.A.W. and N.S.P. acknowledge support received from NSF award number 2607948, which supports the NANOGrav Physics Frontiers Center.

\facilities{Arecibo, GBT}

\software{\texttt{enterprise} \citep{enterprise},
\texttt{enterprise\textunderscore extensions} \citep{ee},
\texttt{PTMCMCSampler} \citep{ptmcmc},
\texttt{hasasia} \citep{hasasia_code},
\texttt{libstempo} \citep{libstempo},
\texttt{tempo} \citep{tempo},
\texttt{tempo2} \citep{tempo2},
\texttt{PINT} \citep{pint},
\texttt{matplotlib} \citep{matplotlib},
\texttt{astropy} \citep{astropy, astropy:2013}
}

\bibliographystyle{aasjournal}
\bibliography{cw}

\appendix
\section{Appendix}\label{sec:appendix}
\begin{table}[h]
    \centering

\begin{tabular}{lll cccc}
\toprule
\textbf{Injection} & \textbf{Parameter} & \textbf{$f_{\rm GW}$} & \textbf{$ \mathrm{log_{10}} \sigma_{\rm min}$}   & \textbf{$\alpha$}  \\
\midrule
\multirow{12}{*}{\textbf{GWB}} 
 & \multirow{3}{*}{$\sigma_h$} 
   & $2/T$ & $-0.65 \pm 0.001$ & $-0.65 \pm 0.001$ \\
 & & $6/T$ & $-0.44 \pm 0.003$ & $-1.25 \pm 0.006$ \\
 & & $7/T$ & $-0.43 \pm 0.003$ & $-1.31 \pm 0.012$ \\
 \cmidrule(lr){2-5}
 & \multirow{3}{*}{$\sigma_f$} 
   & $2/T$ & $-1.46 \pm 0.001$ & $-0.91 \pm 0.001$ \\
 & & $6/T$ & $-1.79 \pm 0.003$ & $-1.41 \pm 0.007$ \\
 & & $7/T$ & $-1.74 \pm 0.006$ & $-1.49 \pm 0.020$ \\
 \cmidrule(lr){2-5}
 & \multirow{3}{*}{$\sigma_m$} 
   & $2/T$ & $-0.54 \pm 0.004$ & $-0.64 \pm 0.003$ \\
 & & $6/T$ & $-0.66 \pm 0.007$ & $-1.38 \pm 0.013$ \\
 & & $7/T$ & $-0.66 \pm 0.010$ & $-1.49 \pm 0.028$ \\
 \cmidrule(lr){2-5}
 & \multirow{3}{*}{$\sigma_\phi$} 
   & $2/T$ & $-0.71 \pm 0.002$ & $-0.85 \pm 0.002$ \\
 & & $6/T$ & $-0.50 \pm 0.005$ & $-1.51 \pm 0.010$ \\
 & & $7/T$ & $-0.35 \pm 0.007$ & $-1.73 \pm 0.023$ \\
\midrule
\multirow{12}{*}{\textbf{No GWB}} 
 & \multirow{3}{*}{$\sigma_h$} 
   & $2/T$ & $-0.49 \pm 0.002$ & $-0.69 \pm 0.001$ \\
 & & $6/T$ & $-0.60 \pm 0.001$ & $-0.92 \pm 0.001$ \\
 & & $7/T$ & $-0.42 \pm 0.001$ & $-0.85 \pm 0.001$ \\
 \cmidrule(lr){2-5}
 & \multirow{3}{*}{$\sigma_f$} 
   & $2/T$ & $-1.45 \pm 0.001$ & $-0.79 \pm 0.000$ \\
 & & $6/T$ & $-2.01 \pm 0.000$ & $-0.96 \pm 0.000$ \\
 & & $7/T$ & $-1.80 \pm 0.001$ & $-0.98 \pm 0.001$ \\
 \cmidrule(lr){2-5}
 & \multirow{3}{*}{$\sigma_m$} 
   & $2/T$ & $-0.40 \pm 0.008$ & $-0.72 \pm 0.004$ \\
 & & $6/T$ & $-0.71 \pm 0.001$ & $-1.12 \pm 0.002$ \\
 & & $7/T$ & $-0.44 \pm 0.002$ & $-1.06 \pm 0.002$ \\
 \cmidrule(lr){2-5}
 & \multirow{3}{*}{$\sigma_\phi$} 
   & $2/T$ & $-0.35 \pm 0.002$ & $-0.94 \pm 0.001$ \\
 & & $6/T$ & $-0.48 \pm 0.001$ & $-1.25 \pm 0.002$ \\
 & & $7/T$ & $-0.59 \pm 0.001$ & $-1.12 \pm 0.001$ \\
\end{tabular}
    \caption{Power law fit parameters for \autoref{fig:tracks}}
    \label{tab:placeholder}
\end{table}

\end{document}